\documentclass[superscriptaddress,groupedaddress,nofootnoteinbib,12pt]{article}  % for review and submission

\usepackage{graphicx}  % needed for figures
\usepackage{dcolumn}   % needed for some tables
\usepackage{bm,amsmath}        % for math
\usepackage{jcappub}

\def\be{\begin{equation}}
\def\ee{\end{equation}}
\def\ba{\begin{eqnarray}}
\def\ea{\end{eqnarray}}
\def\beq{\begin{eqnarray}}
\def\eeq{\end{eqnarray}}

\def\mpl{M_{\rm P}}

\def\d{\mathrm{d}}
\def\p{{\cal P}}

\def\K{{\cal K}}

\def\L*{{\cal L}_*}
\def\L{\mathcal{L}}
\def\({\left(}
\def\){\right)}
\def\ie{{\it i.e. }}

\def\nn{\nonumber}
\def\p{\partial}
\def\mn{_{\mu \nu}}
\def\stu{St\"uckelberg }
\def\N{\mathcal{N}}
\def\p{\partial}
\def\mupn{^\mu_{\ \nu}}
\def\<{\langle}
\def\>{\rangle}

\def\A{\mathcal{A}}
\def\pa {\partial}

\def\cs2{c_{s}^{2}}

 \def\ep{\varepsilon}

 \def\om{\omega}
 \def\Om{\Omega}
 
 \def\p{\partial}

 \def\R{{\cal R}}
 \def\wed{\wedge}
 \def\be   {\begin{equation}}   \def\ee   {\end{equation}}

 \def\ba  {\begin{eqnarray}}   \def\ea  {\end{eqnarray}}

\begin{document}

\title{Deconstructing Dimensions and Massive Gravity}

\author{Claudia de Rham, Andrew Matas,}
\author{and Andrew J.~Tolley}
\affiliation{Department of Physics, Case Western Reserve University, 10900 Euclid Ave, Cleveland, OH 44106, USA}
%\date{\today}

\abstract{We show that the ghost-free models of massive gravity and their multi-graviton extensions follow from considering higher dimensional General Relativity in Einstein-Cartan form on a discrete extra dimension, according to the Dimensional Deconstruction paradigm. We show that Dimensional Deconstruction is equivalent to a truncation of the Kaluza-Klein tower at the nonlinear level. Higher dimensional gravity is not recovered from a lower dimensional multi-graviton theory in the limit of a continuous extra dimension (infinite Kaluza-Klein tower) due to the appearance of a low strong coupling scale that depends on IR physics.
This strong coupling scale, which is associated with the mass of the lowest Kaluza-Klein mode, controls the onset of the Vainshtein mechanism and is crucial to the theoretical and observational viability of the truncated theory.
}

\maketitle

\section{Introduction}

It is a familiar and old idea that a theory of gravity with compactified extra dimensions may be viewed as a four dimensional theory of multiple gravitons, \ie Kaluza-Klein (KK) modes. In its simplest realization, gravity in five dimensions with Planck scale $M_5$ when compactified on a circle of size $R$ gives rise to a four dimensional theory of a single massless and multiple massive gravitons of masses $m_n^2 = (2 \pi n/R)^2$ with a four dimensional Planck mass $M_4^2 =M_5^3 R$. In addition there exists a massless scalar (radion) and massless vector field. The total number of KK modes $N$ is set by the requirement of validity of the effective field theory $m_n \le M_5$, which implies $N \sim M_4^2/M_5^2$. The hierarchy between the four dimensional and five dimensional Planck scales is thus directly tied to the number of species in the theory, \cite{Dvali:2007hz}. Phrased in this way, multi-graviton effective field theories are a generic prediction of many modern proposals for beyond the standard model physics.

An alternative to the KK paradigm was suggested in \cite{ArkaniHamed:2001ca,ArkaniHamed:2001nc} in which one (or more) effective extra dimensions could emerge from a theory of a finite number of massive gauge fields or gravitons living in four dimensions. This idea, coined `Dimensional Deconstruction', can be viewed as taking a five dimensional gauge or gravity theory and discretizing the extra dimension. This approach was effective when applied to gauge theories, however there are novel challenges when applying deconstruction to gravity.  For one thing General Relativity is an intrinsically more complex theory than Yang-Mills, and so while the spectrum of the dimensionally deconstructed theory has been well studied perturbatively, \cite{ArkaniHamed:2002sp,Deffayet:2005yn,Deffayet:2003zk},  a naive discretization fails because
of the breakdown of unitarity (appearance of ghost(s)) at an undesirably low scale, \cite{ArkaniHamed:2003vb,Schwartz:2003vj}.
Some of these issues can be addressed by allowing the extra dimension(s) to be warped. For interesting work along these lines see \cite{Gallicchio:2005mh, Randall:2005me,Deffayet:2004ws,Kiritsis:2008at}, however in this work we shall consider a flat extra dimension. Deconstruction can also be studied in a holographic context where each site is taken to have AdS asymptotics, in this case it may be possible to find other interaction terms that avoid the traditional problems of massive gravity \cite{Kiritsis:2008at}.

The ghosts appearing in the discretization procedure are nothing more than a manifestation of the familiar Boulware-Deser ghost \cite{Boulware:1973my}, (see also \cite{Deffayet:2005ys}). The recent developments in the successful formulation of ghost-free massive theories of gravity and extensions to multi-gravitons, \cite{deRham:2010ik,deRham:2010kj,deRham:2011rn,deRham:2011qq,Hassan:2011hr,Hassan:2011tf,Hassan:2011ea,Hassan:2011zd,Mirbabayi:2011aa,Hassan:2012qv,Deffayet:2012nr} have transformed the situation and it behoves us to reconsider these previous arguments, especially given that previous work with massive gravity \cite{Gabadadze:2009ja,deRham:2009rm} and related theories such as the Galileon and DBI-Galileon \cite{Nicolis:2008in,deRham:2010eu,deRham:2010gu} has shown that there is a connection between massive gravity and extra dimensional gravitational theories that before now has not been fully elucidated.

Given an extra dimensional picture of massive gravity, we are in a better position to understand or explain certain aspects of the dRGT \cite{deRham:2010ik,deRham:2010kj} massive gravity theory and its bigravity \cite{Hassan:2011zd} and multi-gravity \cite{Hinterbichler:2012cn} extensions. For example, we now see that the specific ghost-free form of massive gravity is inherited from the consistency of General Relativity. An extra dimensional interpretation may also shed light on the quantum stability of the theory (see Refs.~\cite{deRham:2012ew,deRham:2013qqa}) and on the presence of superluminal propagation around spherically symmetric backgrounds of Galileon and multi-Galileon theories \cite{Hinterbichler:2009kq,Goon:2010xh,Adams:2006sv,Nicolis:2009qm,Dubovsky:2005xd,deFromont:2013iwa,Garcia-Saenz:2013gya} and in massive gravity itself \cite{Gruzinov:2011sq,deRham:2011pt,Deser:2012qx}, (note however that superluminal propagation does not necessarily imply the existence of stable closed timelike curves \cite{Babichev:2007dw,Burrage:2011cr,deRham:2013hsa}).
Recent work has established that within massive gravity, stability about spherically symmetric solutions requires a change of boundary conditions which in turn removes the superluminal propagation about these configurations, \cite{Berezhiani:2013dw,Berezhiani:2013dca}.

Let us give a simple derivation of the dRGT theory of massive gravity from extra dimensions. Consider the spatial ADM decomposition (Arnowitt, Deser and Misner,  see Ref.~\cite{Arnowitt:1962hi}) of the five dimensional metric in the gauge where the lapse $\N$ is unity and the shift vanishes: $\d s^2 = \d y^2 + g_{\mu \nu}(x,y) \d x^{\mu} \d x^{\nu}$. In this gauge the extrinsic curvature $K_{\mu \nu}$ is given by
\be
K_{\mu \nu} = \frac{1}{2} \partial_y g_{\mu \nu}\,.
\ee
Suppose we now replace the extra dimension by two points $y_1$ and $y_2$ at distance $1/m$ and allow the metric at $y_1$ to remain dynamical $g_{\mu \nu}(y_1)=g_{\mu \nu}$, with the metric at $y_2$ fixed to a reference metric $g_{\mu \nu}(y_2)=f_{\mu\nu}$, (allowing this to be dynamical will simply give us bigravity, \cite{Hassan:2011zd}). In previous work, the $y$-derivative was approximated by $K_{\mu \nu} = m (f_{\mu \nu}- g_{\mu\nu})$. In doing so we inevitably introduce a ghost into the theory, since it is known that gravity with the mass terms build out of a scalar polynomial of $(g_{\mu \nu}- f_{\mu\nu})$ will suffer from the Boulware-Deser ghost, \cite{Boulware:1973my}.

The resolution of this problem is remarkably simple - we should discretize the vielbeins and not the metric! We will then exploit the fact that ghost-free dRGT massive gravity is simple when written in vielbein variables, \cite{Nibbelink:2006sz,Hinterbichler:2012cn,Chamseddine:2011mu}. To see this we denote $g_{\mu\nu}(x,y) = e_{\mu}^{a}(x,y)  e_{\mu}^{b}(x,y) \eta_{ab}$ so that
\be
K_{\mu \nu} = \frac{1}{2} \partial_y g_{\mu\nu} = \frac{1}{2} \left( e_{\mu}^a (\partial_y e_{\mu}^{b}) \eta_{ab} +( \partial_y e_{\mu}^a)   e_{\nu}^{b} \eta_{ab} \right)\,.
\ee
Discretizing the vielbein derivative in the sense $\partial_y e_{\mu}^a \to m (e_{\mu}^{2,a}- e_{\mu}^{1,a})$ then the extrinsic curvature evaluated at $y_1$ associated with the dynamical metric is given by
\ba
K_{\mu\nu} \to \K\mn &=& \frac{m}{2} \left( e_{\mu}^{1 ,a}  (e_{\nu}^{2,b}- e_{\nu}^{1,b} )\eta_{ab}+ e_{\nu}^{1 ,a}  (e_{\mu}^{2,b}- e_{\mu}^{1,b} )\eta_{ab}  \right) \\
 &=& -m \left( g_{\mu \nu} - \frac{1}{2} (e_{\mu}^{1 ,a}  e_{\nu}^{2,b} + e_{\nu}^{1 ,a}  e_{\mu}^{2,b} ) \eta_{ab}\right)\,.
\ea
Although we have fixed the reference metric $f_{\mu\nu}$, the vielbein formalism introduces an additional local Lorentz symmetry. We shall see later that if we use local Lorentz invariance to make the five dimensional  gauge choice $\Omega_y^{AB}=0$ before discretization, and use the vanishing of the fifth component of the torsion, $T^{A}_y=0$, where $\Omega^{AB}$ is the spin connection and the torsion is $T^A=\d e^A+ \Omega^A{}_B \wedge e^B$, then we can infer that the vielbeins satisfy the Deser-van Nieuvenhuizen condition (see \cite{Hoek:1982za,Deffayet:2012zc} for discussion): $e_{\mu}^{1 ,a}  e_{\nu}^{2,b} \eta_{ab }= e_{\nu}^{1 ,a}  e_{\mu}^{2,b} \eta_{ab}$.

Given this fact, we have $\K_{\mu\nu}  = -m \left ( g_{\mu \nu} - e_{\mu}^{1 ,a}  e_{\nu}^{2,b} \eta_{ab} \right)=- m \left ( g_{\mu \nu} - g_{\mu \alpha}e^{\alpha}_{ 1 ,a}  e_{\nu}^{2,a}  \right)$,
\be
e^{\mu}_{ 1 ,a}  e_{\alpha}^{2,a} e^{\alpha}_{ 1 ,b}  e_{\nu}^{2,b} = e^{\mu}_{ 1 ,a}  e_{\alpha,b}^{2} e^{\alpha,a}_{ 1 }  e_{\nu}^{2,b} =g^{\mu \alpha} f_{\alpha \nu}\,,
\ee
and thus we find the following expression for the discretized extrinsic curvature
\be
\K\mupn = -m \left( \delta\mupn -\(\sqrt{g^{-1} f}\)\mupn \right)\,.
\label{Ksqrt}
\ee
In other words, discretizing the extra dimension directly in the vielbein language automatically generates the square root structure characteristic of the dRGT model of massive gravity. It is now straightforward to see that taking the spatial ADM form for the action for five-dimensional gravity, and replacing the extrinsic curvature with the above form, we generate a specific example of dRGT massive gravity. We shall clarify the details of this below but the above argument captures the essence of the derivation.

Furthermore we also establish in this paper the relationship between the truncated KK theory and the deconstruction framework: They are {\it equivalent} under a field redefinition that is essentially a discrete Fourier transform. We may write the mass term in a generalized deconstructed theory as
\be
S_{\rm mass} =\frac{m^2 \mpl^2}{4} \sum_{j, j',j''=1}^N \int\  \ep_{abcd} \alpha_{jj'}\alpha_{jj''} e_j^a \wed e_j^b \wed e_{j'}^c \wed e_{j''}^d\,,
\ee
where $\alpha_{jj'}$ are coefficients determining the discretization procedure, satisfying $\alpha_{jj'}=-\alpha_{j'j}$. We show that truncating a KK tower in a gauge where the lapse is unity is exactly equivalent to the above theory with a specific choice of $\alpha_{jj'}$ coefficients.

Finally, we study the continuum limit of the deconstruction framework. We find that the discretized theory contains a strong coupling scale that prevents us from taking a smooth continuum limit:
\be
\Lambda_c \sim \left( M_5 / R \right)^{1/2} \sim \left( M_4 \, m_1^2 \right)^{1/3},
\ee
where $R$ is the size of the extra dimension, $M_4$ is the four dimensional Planck mass, and $m_1 \sim m/N \sim 2\pi/R$ is the mass of the lowest KK mode.\footnote{Related results were obtained in \cite{ArkaniHamed:2003vb,Schwartz:2003vj} using a metric discretization that led to a ghost via a higher derivative operator at an IR dependent scale. However, in the present case (unlike in  \cite{ArkaniHamed:2003vb,Schwartz:2003vj}) there is no ghost at this scale and as such it is not necessarily the cutoff of the effective theory. It is a strong coupling scale at which perturbation theory breaks down and non-perturbative techniques must be employed. The Vainshtein mechanism allows us to make sense of physics above this scale and this is central to the observational relevance of multi-graviton theories.}

This strong coupling scale $\Lambda_c$ does not depend on the mass $m$ of the highest Kaluza-Klein mode, but rather on the IR scale - the size $R$ of the extra-dimension - through the mass of the lowest Kaluza-Klein mode. In other words, the addition of higher mass KK modes does not raise the strong coupling scale beyond what it was already for a single massive graviton. To recover the continuum limit, one should be able to send $N\to \infty$, while keeping the scale $\Lambda_c$ at or above the five-dimensional Planck scale. This is impossible in this context.
The strong coupling scale $\Lambda_c$ remains below the five-dimensional Planck scale $M_5$ even when we push the mass of the highest mode to the Planck scale, $m=M_5$ (since in fact $\Lambda_c$ is independent of $m$).
We do not interpret this strong coupling scale as a cutoff of an effective field theory, but rather as the onset of the Vainshtein mechanism; nevertheless the existence of this scale clearly prevents a smooth continuum limit to five-dimensional General Relativity.

The previous strong coupling problem is tied to a failure of including the lapse for $y$ diffeomorphisms, $\N$, in our discretization scheme. In the continuum theory we show that the gauge choice $\N=1$ leads to an apparent strong coupling issue, in the sense that the helicity-0 mode of the graviton must be canonically normalized by
\be
\pi\rightarrow \pi/\pa_y\,.
\ee
The origin of this is easy to see. Under an infinitesimal diffeomorphism, $\N^2$ transforms as $\N^2 \rightarrow \N^2 + 2 \N^2 \partial_y \xi^y + \xi^y \pa_y (\N^2)\dots$. To set the gauge $\N = 1$ we must perform a {\it non-local} (in $y$) coordinate transformation which becomes ill-defined for low momentum KK modes with $\pa_y \sim 0$. Classical perturbation theory would break down in this gauge, even in the continuum theory where the lapse is frozen, because $\pa_y$ can be made arbitrarily small. In the continuum theory, this apparent strong coupling is a gauge artifact. However to remove it one needs to use the gauge symmetry associated with the lapse. In discretizing we lose this gauge freedom, and hence the strong coupling becomes physical and not pure gauge.
Since the apparent low strong coupling scale cannot be avoided in the continuum theory if the lapse is fixed, we conjecture that a discretization procedure that keeps the lapse would therefore result in a discretized theory with a smooth continuum limit. However, fixing the gauge $\N=1$ in the continuum theory sets precisely the gauge we need to derive ghost-free multi-gravity theories. Instead we expect theories that include the lapse to propagate new degrees of freedom at the scale $m$, the mass of the highest mode. Thus there is a tension between introducing a low strong coupling scale and obtaining a consistent truncated theory with the correct number of propagating degrees of freedom at the scale of the highest mode. \\

The rest of the paper is organized as follows. In section \ref{sec:vielbein} we properly derive the ghost-free bi-gravity theory and dRGT massive gravity in the vielbein language, and
discuss the extension of these results to an arbitrary number of sites. We then reintroduce gauge invariance via Lorentz and diffeomorphism \stu fields in section~\ref{sec:gauge_invariance}. The mapping between deconstruction and the truncated KK theory is discussed in section~\ref{sec:relation} as well as the coupling with matter. The strong coupling scale is then computed and discussed in section~\ref{sec:StrongCOupling} and we summarize our results as well as present some open avenues in section \ref{sec:Discussion}.

\section{Vielbein Discretization}
\label{sec:vielbein}

\subsection{Five-Dimensional ADM Split}
\label{sec:5d-adm}

It is straightforward to see that we generate a specific example of dRGT massive gravity by taking the spatial ADM form for the action for five-dimensional gravity,
\ba
S^{5d}_{\rm ADM}=\frac{M_5^3}{2}\int \d y \d^4 x \sqrt{-g}\(^{(4)}R[g]+[K]^2-[K^2]\)\,,
\ea
where as mentioned previously, $K\mn$ is the extrinsic curvature along the extra dimension and we use the notation that square brackets represent the trace of a tensor.

As explained above, discretizing the extra dimension in the vielbein language is equivalent to replacing the extrinsic curvature with the above square root function \eqref{Ksqrt} of the metric and the reference metric and simultaneously replacing the integral over the extra dimension with its projection over one site\footnote{Alternatively
one can also consider the sum of the different sites, $\int \d y  \L(x,y)\longrightarrow m^{-1} \sum_{j}\L(x,y_j)$, and obtain instead a theory of multi-gravity with as many interacting and dynamical spin-2 fields as there are sites.},
\ba
\label{replacement1}
\int \d y  \, \L(x,y)\hspace{10pt}&\longrightarrow& \hspace{10pt}m^{-1}\, \L(x,y_1)\\
K\mupn \hspace{10pt}&\longrightarrow& \hspace{10pt} m \,\K\mupn(g,f)\,.
\label{replacement2}
\ea
In the case of two-sites, this leads to a specific four-dimensional theory of massive gravity,
\ba
S^{4d}=\frac{\mpl^2}{2}\int \d^4 x \sqrt{-g}\(^{(4)}R[g]+m^2\([\K]^2-[\K^2]\)\)\,,
\ea
with $\mpl^2=\frac{M_5^3}{m}$.
As we shall see in what follows, by changing the discretization ever so slightly, \ie  by giving a different weight to the different sites,
we can easily generalize the deconstruction procedure to obtain all the possible four-dimensional mass terms
\ba
S^{4d}=\frac{\mpl^2}{2}\int \d^4 x \sqrt{-g}\(^{(4)}R[g]+2 m^2\(\L_2+\alpha_3 \L_3+\alpha_4 \L_4\)\)\,,
\ea
with
\ba
\label{L2}
\L_2&=& \frac{1}{2!}\bar \varepsilon^{\mu\nu\alpha\beta}\bar \varepsilon_{\mu'\nu' \alpha\beta}\K^{\mu'}_\mu \K^{\nu'}_\nu\\
\label{L3}
\L_3&=&\frac{1}{3!}\bar \varepsilon^{\mu\nu\alpha\beta}\bar \varepsilon_{\mu'\nu' \alpha'\beta}\K^{\mu'}_\mu \K^{\nu'}_\nu\K^{\alpha'}_\alpha\\
\label{L4}
\L_4&=&\frac{1}{4!}\bar \varepsilon^{\mu\nu\alpha\beta}\bar \varepsilon_{\mu'\nu' \alpha'\beta'}\K^{\mu'}_\mu \K^{\nu'}_\nu\K^{\alpha'}_\alpha\K^{\beta'}_\beta\,.
\ea
Here $\bar \varepsilon^{\mu\nu\alpha\beta}$ is the Levi-Civita tensor (not symbol, which will be denoted as $\ep$ in what follows, $\ep^{0123}=\ep^{01234}=1$).
However it is hard to see these additional terms in the metric-ADM form, to obtain them we shall switch to vielbein form.

\subsection{Five-Dimensional Einstein-Cartan Formulation}
We now derive properly the mass term by working in the vielbein language and show how the mass term $[\K^2]-[\K]^2$ arises after discretization.
The five dimensional vielbein $E^A_M$ are
\be
g_{MN}=E_M^A E_N^B \eta_{AB},
\ee
where $A,B,\dots$ represent five dimensional local Lorentz indices while $M,N,\dots$ represent spacetime (coordinate) indices. Spacetime indices are raised and lowered with the spacetime metric $g_{MN}$, while local Lorentz indices are raised and lowered with the Minkowski metric $\eta_{AB}$.

Working in the second order form, we can impose the torsionless condition
\be
T^A=\d E^A + \Om^A_{\ \ B}\wedge E^B = 0.
\ee
This condition, along with the antisymmetry of the spin connection $\Om^{AB}=-\Om^{BA}$, uniquely determines the spin connection in terms of the vielbeins
\be
\Om^{AB}_M = \frac{1}{2}E_M^C(O^{AB}_{\ \ \ \ C}-O_C^{\ \ AB}-O^{B\ \ A}_{\ \ C}),
\ee
where $O^{AB}_{\ \ \ \ C}\equiv E^{AM} E^{BN}\(\partial_{M}E_{NC}-\partial_{N}E_{MC}\)$ are the objects of anholonomity. The Riemann curvature 2-form is constructed out of the spin connection in the usual way
\be
\R^{AB}=\d \Om^{AB}+\Om^{A}_{\ \ C}\wed\Om^{CB}\,.
\ee
The five dimensional Einstein-Hilbert term is then given by
\be
S_{\rm EH}^{(5)}=\frac{1}{3!}\frac{M_5^3}{2}\int\ep_{ABCDE}  \, \R^{AB}\wed E^C\wed E^D \wed E^E+S_{\rm bdy},
\ee
where $S_{\rm bdy}$ is the Gibbons-Hawking-York boundary term. The factor $3!$ is a normalization factor needed to convert the action in vielbein language to the action in metric language; in $d$ spacetime dimensions, this factor is $(d-2)!$. The Riemann 2-form in terms of tensor components is $\R^{AB} = \frac{1}{2} \R^{AB}_{\mu\nu} \d x^{\mu} \wedge \d x^{\nu}$.

\subsubsection{Gauge Fixing}

We now perform a standard 4+1 split along the spacelike direction $y$. After using 4 local Lorentz transformations to set $E^5_{\mu}=0$, the vielbeins take their standard ADM form
\be
E^a=e^a+N^a\d y,\ \ \ E^5=\N\d y\,.
\ee
The lapse $\N$ parameterizes the coordinate distance separating nearby hypersurfaces. After discretization we no longer have this reparameterization, so we simply fix one of our diffeomorphism (diff) gauge symmetries with $\N=1$. Meanwhile the shifts $N^a$ parameterize different ways of linking the hypersurfaces together. After discretization, these fields correspond to the \stu fields. We will also use 4 more diffs to pick a gauge with $N^a=0$ and we reintroduce the \stu fields in section~\ref{sec:gauge_invariance}.

Before moving on, one comment is in order with regards to this gauge. As in KK theory we expect for a finite size extra dimension, an additional massless scalar mode (radion) and a massless vector mode. These zero-modes can be accounted for by allowing $\N=\N(x)$ to be independent of $y$, and taking $E^5=\N(x)( \d y+ A_{\mu}(x) \d x^{\mu})$ where $A_\mu(x)$ is also independent of $y$. It is easy to see that this form preserves the $U(1)$ gauge symmetry $y \rightarrow y + \chi(x)$, $A_{\mu}(x) \rightarrow A_{\mu}(x) - \partial_{\mu} \chi(x)$ which confirms the interpretation of $A_{\mu}$ as a $U(1)$ gauge field. For simplicity in the rest of the analysis we will neglect these well understood massless degrees of freedom.

In this gauge the spin connection is given by
\ba
\Om^{ab}&=&\om^{ab}+\Om^{ab}_y\d y, \nn \\
\Om^{5a}&=&K^a\,,
\ea
where the $y$ component of the spin connection is given by
\be
\Om^{ab}_y=\frac{1}{2}\(e^{\mu a} \partial_y e^b_\mu - e^{\mu b}\pa_y e^a_\mu\),
\ee
and the one form $K^a$ is given in terms of the vielbeins by
\be
K^a=\frac{1}{2}\left(e^{\nu b}\pa_y e^a_\nu + e^{\nu a}\pa_y e^b_\nu \right)e_{\mu b} \d x^\mu.
\ee
From this we can see that $K^a$ is related to the extrinsic curvature $K_{\mu\nu}=\frac{1}{2}\partial_y g_{\mu\nu}$ by
\be
K^a_\mu = e^{\nu a} K_{\mu\nu}.
\ee
We use our remaining 6 local Lorentz transformations to fix a gauge where $\Om^{ab}_y=0$. At this point we have used 5 diffs and 10 local Lorentz transformations, so we have completely fixed all of our gauge freedom. The resulting theory of massive gravity we find will then be expressed in unitary gauge, with no \stu fields corresponding to diffeomorphisms or local Lorentz transformations.

In this gauge the Riemann tensor is given by
\ba
\R^{ab}&=&R^{ab}-K^a\wed K^b - \pa_y \omega^{ab} \wed \d y, \nn \\
\R^{5a}&=&\bar{\d} K^a + \om^{a}_{\ \ b}\wed K^b - \pa_y K^a \wed \d y\,,
\ea
where $\bar{\d}$ is the four dimensional exterior derivative (for example the action of $\bar{\d}$ on a scalar $f$ is given by $\bar{\d}f=\pa_\mu f \d x^\mu$), and $R^{ab}=\bar{\d}\om^{ab}+\om^{a}_{\ \ c}\wed\om^{cb}$ is the four dimensional Riemann tensor.
Finally, the Einstein Hilbert action is then finally expressed as
\ba
\label{eq:EH-4+1-split}
S_{\rm EH}^{(5)}=\frac{1}{2!}\frac{M_5^3}{2}\int \ep_{abcd} &\Big(&  R^{ab}\wed e^c \wed e^d - K^a\wed K^b \wed e^c \wed e^d \nn \\
&+& 2K^a\wed \pa_y e^b \wed e^c \wed e^d \Big) \wed \d y\,.
\ea
Here we have chosen to integrate by parts so that there are no $y$ derivatives on $K^a$. This will allow us to avoid discretizing second derivatives in what follows.

\subsubsection{Discretization}

We now discretize along the $y$ direction. In this section we consider the case of 2 sites for simplicity although the multi-sites extension is straightforward. Using the following prescription
\ba
&& \pa_y e^a_\mu \rightarrow m(e_{\mu}^{2,a} - e_{\mu}^{1,a}) \ \ {\rm on\ site\ 1} \nn\\
 && \quad \quad \quad \rightarrow m(e_{\mu}^{1,a}-e_{\mu}^{2,a})\ \ {\rm on\ site\ 2} \nn\\
 && \int f_\mu(x,y) \d x^\mu \wed \d y\rightarrow \frac{1}{m}\sum_{j=1}^2 \int f_{j,\mu}(x) \d x^\mu\,.
\ea
The condition that $\pa_y e^a_\mu$ on site 2 is $m(e_{\mu}^{1,a}-e_{\mu}^{2,a})$ follows from our implicit assumption of periodic boundary conditions $e^{3,a} = e^{1,a}$.
The gauge choice $\Om^{ab}_y=0$ after discretization implies
\be
e^{1,\mu a} e_{\mu}^{2,b} = e^{1,\mu b} e_{\mu}^{2,a},
\ee
which is precisely the Deser-van Nieuvenhuizen condition (`symmetric vierbein' condition). Note that here we have not assumed that this condition is true, it follows in second order form from discretization in the specific five-dimensional gauge we have chosen. As a consequence of this discretization procedure, and the Deser-van Nieuvenhuizen condition, the extrinsic curvature becomes
\be
K^a \rightarrow m(e^{2,a} - e^{1,a})\,.
\ee
Then it is straightforward to see that the discretized five-dimensional Einstein Hilbert action can be written as
\ba
\label{2pts}
S_{\rm EH}^{(5)} \rightarrow \frac{\mpl^2}{4}\int \ep_{abcd}& \Big( R^{1,ab}\wed e^{1,c} \wed e^{1,d} + m^2\A^{abcd}(e^1,e^2)\Big) +(1\leftrightarrow 2),
\ea
where $\mpl^2\equiv M_5^3 / m$ and
with
\ba
\A^{abcd}(e,f)=(f^{a}-e^{a})\wed(f^{a}-e^{a})\wed e^{c} \wed e^{d }\,,
\ea
for any two vielbeins $e$ and $f$. This is a ghost-free bigravity theory, as shown in \cite{Hinterbichler:2012cn}. By rescaling the metric on one site, and taking the decoupling limit where its associated Planck mass is infinite, we recover dRGT massive gravity.\footnote{As shown in \cite{Fasiello:2013woa} the limit of bigravity to massive gravity is a scaling limit. Rather than fixing the dynamics of the $f$-metric, it amounts to decoupling their dynamics by sending the interactions of the second metric to zero. In performing this limit for a generic metric $f$, we may need to subtract from the action an infinite non-dynamical counterterm as in \cite{Fasiello:2013woa}.}

\subsection{Generalized Mass Term}

We can generate a more general mass term by altering our discretization procedure. As is well known from numerical analysis, discretization of a nonlinear theory is not unique. For instance we may choose to generalize our formula for the derivative by including additional sites to improve convergence.
For example, consider this more general discretization
\ba
\pa_y e^a_\mu &\rightarrow& m(\alpha e_{\mu}^{j+1,a} + \beta e_{\mu}^{j,a} - (\alpha+\beta) e_{\mu}^{j-1,a}) \ \ {\rm on\ site\ } j\,.
\ea
with $\beta=1-2\alpha$.

Alternatively, even if we work with the two-site derivative we can choose to replace nonlinear products by combinations at different sites, for instance in the expression
\be
\L\supset\int \d y \, e^a(x) \wedge e^b(x) \wedge \pa_y e^c \wedge \pa_y e^d\,,
\ee
$e^a(x^j)$ can be replaced with $(r_j e^{j+1,a}+ (1-r_j) e^{j,a})$ with two free parameters for each site $j$ so as to give
\be
\L\supset\sum_j  (r_j e^{j+1,a}+ (1-r_j) e^{j,a}) \wedge (s_j e^{j+1,b}+ (1-s_j) e^{j,b}) \wedge (e^{j+1,c}- e^{j,c}) \wedge (e^{j+1,d}- e^{j,d})  \,,\nn
\ee
with $0 \le r_j \le 1$, $0 \le s_j \le 1$.With this choice, the gauge choice $\Om^{ab}_y=0$ is still equivalent to the Deser-van Nieuvenhuizen gauge after discretization and the extrinsic curvature is still
\be
K^a \rightarrow e^{2,a} - e^{1,a}.
\ee
Then the discretized action is the same as in \eqref{2pts} with the more general 2-parameter family mass  term,
\ba
\label{2pts_rs}
S_{\rm EH}^{(5)} \rightarrow \frac{\mpl^2}{4}\int \ep _{abcd}& \Big( R^{1,ab}\wed e^{1,c} \wed e^{1,d} + m^2\A_{r,s}^{abcd}(e^1,e^2)\Big) +(1\leftrightarrow 2),
\ea
with
\ba
\label{def_A}
\A_{r,s}^{abcd}(e,f)=(f^{a}-e^{a})\wed(f^{b}-e^{b})\wed (r e^{c}+(1-r)f^{c}) \wed (s e^{d}+(1-s)f^{d})\,,
\ea
for any two vielbeins $e$ and $f$. $\A$ can be expanded of the form
\ba
\label{def_A}
\A^{abcd}&=&c_0 e^a\wedge e^b \wedge e^c \wedge e^d+c_1 e^a\wedge e^b \wedge e^c \wedge f^d\\
&+&c_2 e^a\wedge e^b \wedge f^c \wedge f^d+c_3 e^a\wedge f^b \wedge f^c \wedge f^d
+c_4 f^a\wedge f^b \wedge f^c \wedge f^d\,,\nn
\ea
with $c_0=rs$, $c_1=(r+s-4rs)$, $c_2=(1-3s-3r+6 rs)$, $c_3=(-2+3s+3r-4 rs)$ and $c_4=(1-s)(1-r)$. This corresponds to the most general potential which, by construction, bears no cosmological constant nor tadpole and is thus a combination of $\L_2$, $\L_3$ and $\L_4$ as expressed in \eqref{L2}, \eqref{L3}, \eqref{L4}.
So this method can generate the most general dRGT or bigravity theory by changing the discretization prescription using only the two-site derivative.

\subsection{Multi-Gravity}
\label{sec:multi-gravity}

We can easily extend our formalism to allow for $N$ sites, generating a multi-gravity theory. The vielbein on the first site is $e^{a, 1}_\mu$ and we use periodic boundary conditions, $e^{a, j+N}_\mu=e^{a, j}_\mu$.
We maintain locality in the auxiliary dimension by demanding the derivative to couple only neighbouring sites after discretization. The discretization procedure is then straightforward
\ba
&& \pa_y e^a_\mu \  \rightarrow \ m(e_{\mu}^{j+1,a} - e_{\mu}^{j,a}) \ \ {\rm on\ site\ } j \nn\\
 && \int f_\mu(x,y) \d x^\mu \wedge \d y\, \rightarrow \ \frac{1}{m}\sum_{j=1}^N \int f_{j,\mu}(x) \d x^\mu\,.
\ea
If the extra dimensions have a boundary, care must be taken to define the derivative on the boundary sites. However we will suppose, in line with the usual KK logic and the original deconstruction proposal, that the auxiliary dimension is compact. Then we may avoid this issue by imposing periodicity in $j$, so that
\be
e_\mu^{j+N,a}=e_\mu^{j,a}.
\ee
The gauge choice $\Om^{ab}_y=0$ after discretization implies
\be
e^{j,\mu a} e_{\mu}^{j+1,b} = e^{j,\mu b} e_{\mu}^{j+1,a}\,.
\ee
As a consequence of this discretization procedure, the above condition, the extrinsic curvature on site $j$ becomes
\be
K_j^a \rightarrow m(e^{j+1,a} -e^{j,a}).
\ee
The discretized action is then
\ba
S_{\rm EH}^{(5)} \rightarrow \frac{\mpl^2}{4} \sum_{j=1}^{N} \int \ep _{abcd} \Big[ R^{j,ab}\wed e^{j,c} \wed e^{j,d} + m^2 \A_{r_j,s_j}^{abcd}(e^j,e^{j+1})\Big]\,,
\ea
where $\A_{r,s}^{abcd}(e,f)$ is defined in \eqref{def_A} for any two vielbeins $e$ and $f$ and any two free parameters $r$ and $s$.
This has the form of a multi-gravity theory as discussed in \cite{Hinterbichler:2012cn}. More general interactions between the multiple vielbein fields and not only the closest neighbors can be obtained by generalizing the discretization procedure to involve more sites.
Since each of the metrics has a contribution from the zero mode, the four dimensional Planck mass seen by the analogue of the KK mode is given by $M_4^2 = N \mpl^2$.

\section{Recovering Gauge Invariance}
\label{sec:gauge_invariance}

\subsection{Linking Fields}
Ordinary four dimensional gravity exhibits both diffeomorphism invariance and local Lorentz invariance which acts on the vierbein indices. In multi-gravity with $N$ gravitons, the $N$ copies of diff and $N$ copies of local Lorentz are broken down to a single copy of each, for which all the vierbeins transform in the same way. We can however easily reintroduce the broken $ 2 \times (N-1) $ symmetries by means of linking or \stu fields. In the present case we need \stu fields for both the broken diff symmetry, which we denote as scalars $\Phi^a_j$ and \stu fields for the broken local Lorentz transformations which we denote as $\Lambda^a{}_{b, i}$ where $\Lambda^a{}_{b, i}$ satisfies $\Lambda^a{}_{b, i} \Lambda^c{}_{d, i} \eta_{ac}= \eta_{bd}$. The idea is to replace the vierbein $e^{j+1,c}$ with one which is covariant under diff and Lorentz transformations at the site $j$, so that the difference $E^{j+1,a}-e^{j,a}$ is covariant at the site $j$. The relevant expression is \cite{ArkaniHamed:2002sp,Nibbelink:2006sz,Ondo:2013wka,Gabadadze:2013ria,Fasiello:2013woa}
\be
E^{j+1,a}_{\mu}(x^{\alpha}) = \partial_{\mu} \Phi_j ^{\beta}(x) \, \Lambda^a{}_{b, j}(x) e^{j+1,b}_{\beta} ( \Phi_j^{\alpha}(x))\,.
\ee
This double \stu trick was used in \cite{Ondo:2013wka} (see also \cite{Gabadadze:2013ria}) to derive the complete decoupling limit of massive gravity including vector modes, and \cite{Fasiello:2013woa} to derive the complete decoupling limit of bigravity models.

The fully covariant multi-gravity action which uses only the nearest site discretization is then
\ba
S_{\rm EH}^{(5)} \rightarrow \frac{\mpl^2}{4} \sum_{j=1}^N \int \ep_{abcd} \Big( R^{j,ab}\wed e^{j,c} \wed e^{j,d}
+m^2 \A_{r_j,s_j}^{abcd}(e^j,E^{j+1})\Big]\,,
\ea
where $\A$ is again given in \eqref{def_A}.
Introduced in this way, every individual term in the sum coming from the site $j$ is invariant under an independent copy of diffs and Lorentz at the site $j$.

\subsection{Relation with Wilson Line Formalism}
In this section we will show how to discretize five-dimensional General Relativity without fixing a gauge, showing a connection between the shift $N_\mu$ and the \stu fields. We use the formalism developed by Refs.~\cite{Deffayet:2005yn,Deffayet:2003zk}.
Rather than simply discretizing the $y$ derivative, we discretize the covariant $y$ derivative
\be
D_y = \pa_y - N^\mu \pa_\mu\,,
\ee
using the prescription
\be
\mathcal{L}_{D_y} T_{\mu_1 \cdots \mu_n} \rightarrow m\left(\hat W_{j, j+1}T_{\mu_1\cdots \mu_n}(x_j)-T_{\mu_1\cdots \mu_n}(x)\right)\,.
\ee
The operators $\hat W_{j, j'}$, closely related to Wilson lines from Yang-Mills theory, are given by
\be
\hat W_{j, j'}=\mathcal{P} e^{\int_{x_j}^{x_{j'}} \d z \mathcal{L}_{D_y}}\,,
\ee
where $\mathcal{P}$ is the path ordering symbol and $\mathcal{L}_{D_y}$ is the Lie derivative with respect to the vector $D_y$. These operators allow us to map tensors on site $j+1$ to tensors on site $j$ in the following sense
\be
\hat W_{j, j+1}T_{\mu_1\cdots \mu_n}(x_{j+1}) = T_{\nu_1\cdots \nu_n}(\Phi_{j,j+1}) \pa_{\mu_1} \Phi^{\nu_1}_{j,j+1} \cdots \pa_{\mu_n} \Phi^{\nu_n}_{j,j+1}\,.
\ee
Thus these operators play the role of introducing the \stu fields $\Phi$ in the continuum theory and amount to a formalization of the \stu trick. The extrinsic curvature is
\be
K_{\mu\nu} = \frac{1}{2\N} \mathcal{L}_{D_y} g_{\mu\nu} = \frac{1}{2\N}\eta_{ab}\left(e^a_\mu \mathcal{L}_{D_y} e^b_\nu + e^b_\nu \mathcal{L}_{D_y} e^a_\mu  \right)\,,
\ee
and discretizing we find
\ba
K_{j,\mu\nu}&=&\frac{m}{2\N_j} \eta_{ab} \left( e_{j,\mu}^a (E_{j,j+1,\nu}^b-e_{j,\nu}^b) - e_{j,\nu}^b(E_{j,j+1,\mu}^a-e_{j,\mu}^a)\right) \, ,\nn \\
K_{j,\nu}^{\mu} &=& -\frac{m}{\N_j} \left(\delta_{\nu}^{\mu}-{\sqrt{g_j^{\mu \alpha} g_{j+1,\alpha \nu}}} \right)\,.
\ea
If we fix the gauge $\N=1$ this is exactly the form of $K_{j,\mu\nu}$ we find by following the \stu procedure from the previous section.

\section{Relationship between Kaluza-Klein and Deconstruction}
\label{sec:relation}

\subsection{Kaluza-Klein Decomposition}

At the linear level it is well known that the deconstruction map is essentially a discrete Fourier transform of the KK map.
Decomposing $g_{\mu \nu} = \eta_{\mu \nu}+ \mpl^{-1} h_{\mu\nu}$ then the relation is given by, \cite{Deffayet:2005yn}
\be
\tilde{h}_{\mu \nu ,n}=\frac{1}{\sqrt{N}}\sum_{j=1}^{N}h_{\mu \nu , j} e^{i \frac{2\pi n}{N} j}\,.
\ee
At the nonlinear level this map is nontrivial in terms of the metric precisely because as we have seen discretizing the metric fails to give the correct mass terms. However, if we perform the KK decomposition in the vierbein language in the gauge we have chosen, then even at the nonlinear level the two prescriptions are exactly discrete Fourier transforms of each other. Thus we may define the map between the nonlinear KK modes and the vierbeins at a given site as
\be
\tilde{e}_{\mu \nu, n}^a=\frac{1}{\sqrt{N}}\sum_{j=1}^{N}e_{\mu \nu, j}^a e^{i \frac{2\pi n}{N} j}\,,
\ee
with the inverse map as
\be
e_{\mu \nu, j}^a = \frac{1}{\sqrt{N}} \sum_{n=-M}^{M} \tilde{e}_{\mu \nu, n}^a e^{-i \frac{2\pi n}{N} j}\,,
\ee
with $M=(N-1)/2$.

These are exact statements, not made in reference to perturbations around any background. Of course the $\tilde{e}_{\mu \nu, n}^a$ do correspond to the mass eigenstates expanded around Minkowski spacetime but not necessarily around a more general background, nevertheless we may choose to use them as field variables non-perturbatively. Since the Fourier transform is nothing more than an invertible field redefinition in the multi-field space, and since field redefinitions do not change the physics, in the end the whole deconstruction framework is equivalent after a field redefinition to the KK framework, provided that both are performed in terms of the vierbein in the gauge we have chosen.

Let us see how this works explicitly. We begin in the first order five-dimensional continuum Einstein-Cartan formalism, \ie working with the spin connection as an independent variable, working with the gauge choice that
\be
E^a = e^a = e^a_{\mu} \d x^{\mu} \quad \text{and} \quad \, E^5 = \d y\,,
\ee
and in addition we choose the same gauge as before for the connection $\Omega^{ab}_y=0$.
With this gauge choice the five-dimensional continuum action
\be
S_{\rm EH}^{(5)}=\frac{M_5^3}{12}\int\ep_{ABCDE}  \, \R^{AB}\wed E^C\wed E^D \wed E^E,
\ee
has the same form as the second order action of Eq.~\eqref{eq:EH-4+1-split},
\ba
S_{\rm EH}^{(5)}=\frac{M_5^3}{4}\int \ep_{abcd} &\Big(&  R^{ab}\wed e^c \wed e^d - K^a\wed K^b \wed e^c \wed e^d \nn \\
&+& 2K^a\wed  e^b \wed e^c \wed \pa_y e^d \Big) \wed \d y\,,
\ea
where we use the same notation as section \ref{sec:vielbein} except that now $\omega^{ab}$ and $K^a$ should all be viewed as independent variables from the vierbein $e^a$.

The idea now it to assume that $y$ is compactified on a circle of radius $R$ so that all the fields may be decomposed into a spectrum of KK modes on that circle in the sense for any function $T(y)= \{ e^a, K^a, \omega^{ab} \}$ we have
\ba
&& T(y) = \frac{1}{\sqrt{N}} \sum_{n =- \infty}^{\infty}  \tilde{T}_n e^{-i \frac{2\pi n y}{R}}  \,  ,
\ea
where the factor of $1/\sqrt{N}$ is introduced for future convenience.
We then take these expressions, substitute into the five dimensional action and integrate over $y$. The integral will impose momentum conservation in the fifth dimension. At this point we truncate the sum over KK modes so that it goes from $n = -M$ to $n= M$.
We then define, given the truncated spectrum for any $T= \{ e^a, K^a, \omega^{ab} \}$
\ba
&& T_j = \frac{1}{\sqrt{N}}\sum_{n =- M}^{M}  \tilde{T}_{n} e^{-i \frac{2\pi n j}{N}}  \, ,
\ea
and the associated inverses
\ba
&& \tilde T_n= \frac{1}{\sqrt{N}}\sum_{n =1}^{N}  {T}_{j} e^{i \frac{2\pi n j}{N}}  \, .
\ea

\subsection{Truncated Kaluza-Klein versus Discretization}

\subsubsection{Ultralocal Operators}
Any expression which was ultralocal in $y$ (independent of $y$ derivatives) in the original expression, maps into an expression which is ultralocal in the sum over $j$. To see this consider an example term
\be
I=\int \d y \, A(y) B(y)
\ee
We first express this into KK modes, and integrate over $y$,
\be
I=\frac{R}{N} \sum_{n = -\infty}^{\infty} A_{-n}  B_n \, .
\ee
We then truncate the spectrum
\be
I_M=\frac{R}{N} \sum_{n = -M}^{M} A_{-n}  B_n \, ,
\ee
and reexpress the truncated KK modes in terms of the discretized expressions
\be
I_M=\frac{R}{N^2} \sum_{n = -M}^{M} \sum_{j=1}^N \sum_{j'=1}^N A_j B_{j'} e^{i \frac{2 \pi n}{N}(j-j')}\,.
\ee
Performing the sum over $n$ gives a Kronecker delta $N \delta_{jj'}$ which in turn gives
\be
I_M=\int \d y  \, A(y) B(y) \rightarrow \frac{R}{N} \sum_{j=1}^N A_j B_j\,,
\ee
which is nothing else but the naive discretized expression for the original integral. It is easy to see that this argument generalizes to arbitrary ultralocal products.
Thus it is only necessary to worry about the terms with $\partial_y $ derivatives.

\subsubsection{Map between Derivative Terms}

To explain how these map through terms that contain derivatives, consider an example term
\be
I=\int \d y \, A(y) \partial_ y B(y)
\ee
We first express this into KK modes,
\be
I=\frac{ R}{N} \sum_{n = -\infty}^{\infty} A_{-n} \frac{2 \pi i }{R} n B_n \, ,
\ee
and then truncate the KK spectrum before reexpressing it in terms of the sites as preformed previously to give
\be
I_M=\frac{R}{N^2} \sum_{n = -M}^{M} \sum_{j=1}^N \sum_{j'=1}^N A_j B_{j'}  \frac{2 \pi i n}{R} e^{i \frac{2 \pi n}{N}(j-j')}\,.
\ee
We now use the fact that
\be
\label{alphadef}
\sum_{n = -M}^{M}  i n \, e^{i \frac{2 \pi n}{N}(j-j')} = (-1)^{j-j'} \frac{N}{2 \sin ((j-j') \pi/N)}= \frac{N^2}{2\pi} \alpha_{jj'}\, .
\ee
Thus finally we have
\be
\int \d y A(y) \partial_ y B(y) \rightarrow \sum_{j=1}^{N} \sum_{j'=1}^N \alpha_{jj'} A_j B_{j'}\,.
 \ee
 This is easily generalized to higher order products
 \be
 \int \d y A(y) C(y) D(y)  \partial_ y B(y) \rightarrow \sum_{j=1}^{N} \sum_{j'=1}^N \alpha_{jj'} A_j C_j D_j B_{j'}\,.
 \ee

\subsubsection{Map for the Five-Dimensional Action}

We now have enough information to determine the form of the five-dimensional Lagrangian after this procedure. Specifically it is
\ba
S_{\rm KK}&=&\frac{M_5^3 R}{4 N} \sum_{j=1}^N  \int  \ep_{abcd} \, \Big( e_j^a \wedge e_j^b \wedge R_j^{cd} - e_j^a \wedge e_j^b \wedge K_j^c \wedge K_j^d \nn \\
&+&  2 m \sum_{j'=1}^N \alpha_{jj'}\ K_{j}^a \wedge e_j^b \wedge e_j^c \wedge e_{j'}^d  \Big) \, ,
\ea
where as before $m=N/R$. We can show that this is equivalent to a multi-graviton theory with a specific choice for the mass terms.
$K^a_{\mu j}$ can be inferred by varying the Lagrangian with respect to $K$,
\be
K^a_j  =m \sum_{j'=1}^N \alpha_{j j'} e^a_{j'} \, .
\ee
Putting this together we see that the truncated KK action is equivalent to a specific multi-graviton action expressed in the vierbein form as
\ba
S_{\rm KK}=\frac{\mpl^2}{4} \sum_j  \int  \ep_{abcd} \, \left( e_j^a \wedge e_j^b \wedge R_j^{cd} + m^2 \sum_{j', j''}\, \alpha_{j j'}\alpha_{j j''} \, e_j^a \wedge e_j^b \wedge  e_{j'}^c \wedge e_{j''}^d  \right) .
\ea
The last thing missing in this formalism are the vector and scalar (radion) zero modes. These are easily introduced by following the standard KK procedure. For instance the radion can be simply accounted for by including $E^5 = e^{\phi(x)/M_4} \d y$ so that the action becomes
\ba
S_{\rm KK}=\frac{\mpl^2}{4} \sum_j  \int  \ep_{abcd} \, \left( e^{\phi/M_4} \, e_j^a \wedge e_j^b \wedge R_j^{cd} + e^{-\phi/M_4} \, m^2 \sum_{j', j''}\, \alpha_{j j'}\alpha_{j j''} \, e_j^a \wedge e_j^b \wedge e_{j'}^c \wedge e_{j''}^d  \right).\nn
\ea
Since the radion $\phi(x)$ is independent of $y$ it commutes with the Fourier transform procedures. The zero-mode vectors are similarly introduced by switching on a $y$-independent $E^5_{\mu}(x)$.

We see that the only difference between Kaluza-Klein theory and the two-site derivative deconstruction method introduced earlier is the precise form of $\alpha_{jj'}$.
In the two-site case we have $\alpha_{jj'} = \delta_{j+1 j'}-\delta_{jj'}$. Although this looks different than the form given in Eq.~(\ref{alphadef}), their Fourier transform is identical at low $n$, reflecting the fact that the two-site discretization has the same KK mass spectrum for modes with $n \ll N$.

Thus we see that the KK framework is {\bf equivalent} to a specific deconstruction framework, with a more sophisticated discretization of the derivative. What we have gained in introducing the two-site deconstruction framework is
\begin{enumerate}
\item A simpler expression for the nonlinear action (namely one that includes only nearest site interactions).
\item A clearer picture of how to truncate the spectrum to a finite number of graviton modes.
\item A clearer picture of how to decouple either the massless or the massive modes by rescaling the associated Planck scales appropriately.
\end{enumerate}

\subsection{Coupling to Matter}
\label{sec:Matter}
If we take the deconstruction paradigm to its extreme, then it is natural to imagine that every matter field comes in $N$ copies which are coupled to the $N$ distinct vierbeins. However, in order to recover five-dimensional Lorentz invariance in the continuum limit we must add a gradient energy term $(\partial_y \rm matter)^2$. This necessarily entails that the matter fields located on different sites admit at least neighbouring site interactions. For instance, for a scalar field $\chi(x,y)$ with a potential $V(\chi(x,y))$ it is natural to define the derivative through the two-site prescription
\be
\partial_y \chi(x,y_j) \rightarrow m (\chi^{j+1}(x)- \chi^j(x))\,,
\ee
so that the matter Lagrangian becomes after discretization
\be
S_{\rm matter} = \frac{1}{m}\int \d^4 x \sum_j  \sqrt{-g^j} \left( - \frac{1}{2} g^{\mu \nu }_j\partial_{\mu} \chi^j \partial_{\nu} \chi^j  - \frac{1}{2} m^2 (\chi^{j+1}-\chi^j)^2 -V(\chi^j)\right)\,,
\ee
where $g^j_{\mu\nu}$ is the metric at site $j$ built out of the vierbein. Since we are working with a vielbein formalism it is straightforward to extend this to fermion fields and additional gauge or $p$-form fields. The current proof of the absence of ghosts in bigravity and multi-vierbein theories do not fully account for interactions of matter fields between different vierbeins. The above interaction appears to be safe because the Hamiltonian will remain linear in the various lapses. The same is not true if we have kinetic interactions between two different sites. It is clearly an interesting question to explore the most general form of ghost-free matter interaction between different sites.

\section{Strong coupling scale and the Recovery of the Fifth dimension}
\label{sec:StrongCOupling}

To derive the strong coupling scale we perform a decoupling limit expansion in the metric language. Integrating out the $(N-1)$ Lorentz \stu fields $\Lambda^a_{b, j}$ we return to the metric language with the symmetrized vierbein expressed in terms of square root combinations of the metric. We then decompose the diff \stu fields as $\Phi^a_j=x^a+\pi^a_j$ where $\pi^a_j$ are the Goldstone bosons associated with the broken diff. We further split the \stu fields into the vector and scalar mode
\ba
\pi^a_j=\frac{1}{m \mpl}B^a_j+\frac{1}{m^2 \mpl}\p^a \pi_j\,,
\ea
and work with the canonically normalized metric perturbation,
\ba
g\mn^j=e^{j, a}_\mu e^{j, b}_\nu\eta_{ab}=\eta\mn+\frac{1}{\mpl}h^j\mn\,.
\ea
The resulting four-dimensional language is then symbolically of the form (for the simplest case of the $[K]^2-[K^2]$ model),
\ba
\label{L4_j}
\L=\sum_{j=1}^N &\Big[&(\p h_j)^2 +m^2 \(h_{j}(x)-h_{j+1}(\Phi^a_j(x))\)^2 \\
&&+ \(h_{j}(x)-h_{j+1}(\Phi^a_j(x))\)\(\p^2 \pi_j(x) + \frac{1}{\Lambda^3}(\p^2 \pi_j(x))^2\)\nn \\
&&+
\sum_{k=0}^\infty (\p B_j(x))^2 \(\frac{\p^2 \pi_j(x)}{\Lambda^3}\)^k+\cdots\Big]\,,\nn
\ea
where the ellipses represent operators that are suppressed by an energy scale larger than $\Lambda=(m^2 \mpl)^{1/3}=(m M_5)^{1/2}$. The vector-scalar interactions in the last line represent the ones found in Refs.~\cite{Ondo:2013wka,Fasiello:2013woa} (see also \cite{Gabadadze:2013ria}). Furthermore additional interactions are hidden within the argument of $h_{j+1}(\Phi^a_j(x))$, as is already manifest in the decoupling limit of bigravity\footnote{See also Refs.~\cite{deRham:2013hsa,Curtright:2012gx} for insight on how to deal with these interactions.}, \cite{Fasiello:2013woa}, with
\ba
\label{Taylor}
h_{j+1}(\Phi^a_j(x))&=&\sum_{k=0}^\infty \p^k h_{j+1}(x) \(\frac{\p \pi_j}{\Lambda^3}\)^k+\cdots \\
&\sim& h_{j+1}(x) \sum_{k=0}^\infty  \(\frac{\p^2 \pi_j}{\Lambda^3}\)^k+\cdots\,,
\label{taylor2}
\ea
after integrations by parts.

\subsection{Strong Coupling Scale}
To find the lowest-energy interaction scale, it is more convenient to follow the behavior of the mass eigenvalues following the prescription introduced in \cite{ArkaniHamed:2003vb,Schwartz:2003vj,Deffayet:2005yn}. Out of the real space quantities $T_j=\{h_j, \pi_j, B_j\}$ defined at the sites $j$, we can define the discrete Fourier transforms
\ba
\tilde{T}_n=\frac{1}{\sqrt{N}}\sum_{j=1}^{N}T_j\ e^{i \frac{2\pi n}{N} j}\,,
\ea
for $n=0,\cdots,N$ and $T_{-n}=T^*_{n}$.
The multi-graviton action then becomes (symbolically),
\ba
\L=\sum_{n=0}^{N-1}\Big[|\p \tilde h_n|^2+m_n^2|\tilde h_n|^2 + \frac{m_n}{m} |\tilde h_n \p^2 \tilde \pi_n | + |\p \tilde B_n|^2
\Big]+\mathcal{L}_{\rm int}
\,,
\ea
where $m_n = m \sin(n/N)$ are the mass eigenstates, $m_n\sim n/N$ for the lowest modes. Note that due to the square root normalization in the discrete Fourier transform, we can see that the four dimensional Planck mass seen by the KK zero mode $n=0$ is given by
\ba
M_4^2 = N \mpl^2 = \frac{N M_5^3}{m} = M_5^3 R\,,
\ea
which is consistent with the expectation from KK theory. In what follows we work with the canonically normalized scalar mode,
\ba
\hat \pi_n= \frac{m_n}{m}\tilde \pi_n\sim \frac{n}{N}\tilde \pi_n\,.
\ea
 For a large number of sites, the interactions that come at the lowest energy scale are the ones arising from the second line of \eqref{L4_j}, using the expansion \eqref{taylor2},
\ba
\mathcal{L}_{\rm int} &\supset& \sum_{j}h_{j+1}(x)\sum_{k=2}^\infty \frac{(\p^2 \pi_j)^{k}}{\Lambda^{3(k-1)}}\\
&=&\sum_{k=2}^\infty\ \sum_{n_1,\cdots,n_k=1}^M \frac{N^{(k+1)/2}}{\Lambda^{3(k-1)}}\tilde h_{-(n_1+\cdots+n_k)} \frac{(\p^2 \hat \pi_{n_1})}{n_1}\cdots \frac{(\p^2 \hat \pi_{n_k})}{n_k}\,,
\label{h ddpi}
\ea
where as before $M=(N-1)/2$. So we see that the cubic interaction $h^*_{2} (\p^2 \pi_1)^2$ is the one that arises at the lowest energy scale, $\Lambda_c=\Lambda/\sqrt{N}=\sqrt{M_5/R}= \(M_4 m_1^2 \)^{1/3}$, and no other interactions arise at such a low scale.
Here $m_1 \sim m/N \sim 1/R$ is the mass of the lowest KK mode. This scale is precisely the strong coupling scale in a single massive graviton theory whose mass is the lowest KK mode.

This vertex contributes to the $\pi_1 \pi_1 \to \pi_1 \pi_1$ scattering amplitude, with a factor $N^3/\Lambda^6$. Other quartic interactions contribute to that same scattering amplitude, but among the quartic ones, the leading contribution goes as $N^2/\Lambda^6$ and can therefore never cancel the contribution going as $N^3/\Lambda^6$. As a result, one already hits strong coupling at the scale of the cubic interaction in \eqref{h ddpi}.

The fact that the IR scale $R$, which determines the size of the fifth dimension does enter the strong coupling scale $\Lambda_c$ (via $N$ dependence), as already shown in \cite{ArkaniHamed:2003vb,Schwartz:2003vj}, implies that in the continuum limit for which $m \sim M_5 \sim \Lambda$, all the lowest graviton modes become strongly coupled below the five-dimensional Planck scale. Note that these conclusions follow from looking at the mass spectrum of modes with $n\ll N$ and are independent of the specific choice of discretization coefficients $\alpha_{jj'}$. In particular the low cutoff will arise in a truncated Kaluza Klein discretization as well as the two-site discretization.

\subsection{General Interactions}

In general, there is no choice of coefficients which would eliminate the interactions found previously using the discretization procedure we have chosen, which is the only one that preserves $5N$ degrees of freedom, \ie the correct number of degrees of freedom for one massless spin-2 field interacting with $(N-1)$ massive spin-2 fields. Nevertheless, even if this had been possible, the interactions with the vectors would always bring back an IR-dependent strong coupling scale.
Considering the most general kind of interactions,
\ba
\L_{\rm int}&\supset& m^2 \mpl^2 \sum_j \left(\frac{h_j}{\mpl}\right)^p \left( \frac{\partial B_j}{m \mpl} \right)^q \left(\frac{\partial^2 \pi_j}{m^2 \mpl} \right)^k \\
&\sim& \frac{1}{N^{\frac{p+q-k-2}{2}}\mpl^{p+q+k-2}m^{q+2k-2}}
\sum_{n,n',n''\cdots} (\tilde h_n)^p (\partial \tilde B_{n'})^q (\p^2 \hat \pi_{n''})^k\,,
\ea
for arbitrary positive powers $p,q,k$. The related scale for such interactions is
\be
M_c = \(N^{(p+q-k-2)/2} \mpl^{p+q-2}m^{q-2}\Lambda^{3 k}\)^{1/(p+2q+3k-4)}.
\ee
We recover the same result as previously where the lowest energy scale arises for $p=1, q=0$ and $k=2$, for which $M_{c,\, {\rm min}}=\Lambda_c=\Lambda/\sqrt{N}$.

Furthermore, we can see that all of the vector-vector-scalar interactions ($p=0,q=2$) come in at the scale $M_c=N^{-1/6}\Lambda$. In addition we see that interactions with large $k$ also scale as $N^{-1/6}$. So there are an infinite number of interactions that come in arbitrarily close to the scale $M_c=N^{-1/6}\Lambda$.

Discretizing the way we have done, there seems to be no way out of the IR dependent strong coupling scale. From a four-dimensional viewpoint this is not necessarily a bad thing as this low strong coupling scale is precisely what allows for a Vainshtein mechanism and the recovery of four-dimensional gravity at high energy. If on the other hand one would like to recover five-dimensional General Relativity in the limit $N\to \infty$  this discretization seems inappropriate, and we will see in what follows how keeping the lapse seems to be required to recover five-dimensional GR in the continuum limit.

\subsection{Continuum Theory}

\subsubsection{Freezing the Lapse}

The origin of the strong coupling scale in the discretized theory, and a possible resolution, can be seen in the continuum theory. Setting $\N=1$ and $N_\mu=0$ we find for the continuum theory that
\ba
\mathcal{L}_{\rm GR, 5d} &=& -\frac{1}{4} {}^{(5)} h^{AB}  {}^{(5)}\mathcal{E}_{ABCD} {}^{(5)}h^{CD} + \mathcal{O}\left({M_5^{-1}}\right) \\ \nn
&=& -\frac{1}{4} h^{\mu\nu} \mathcal{E}_{\mu\nu\rho\sigma} h^{\rho\sigma} -\frac{1}{8}  \left(\left[(\pa_y h)^2\right]-[\pa_y h]^2\right) +\mathcal{O}\left({M_5^{-1}}\right) .
\ea
where ${}^{(5)}\mathcal{E}_{ABCD} = -\frac{1}{2} \Box_5 + \dots$ is the five dimensional Lichnerowicz operator and $ \mathcal{E}_{\mu\nu\rho\sigma} =-\frac 12 \Box+ \dots$ is the four-dimensional counterpart.
Then we introduce the ``scalar part of the \stu field" by doing a linear diff with gauge parameter $\xi^{\mu} =\partial^{\mu} \pi$ so that $h_{\mu\nu} \rightarrow h_{\mu \nu} +2 \Pi_{\mu\nu}$ with $\Pi_{\mu\nu}=\pa_\mu\pa_\nu \pi$:
\be
\mathcal{L}_{\rm GR, 5d}=-\frac{1}{4}h^{\mu\nu} \mathcal{E}_{\mu\nu\rho\sigma} h^{\rho\sigma}  -\frac{1}{8}  \left(\left[(\pa_y h)^2\right]-[\pa_y h]^2\right) -\frac{1}{2} \pa_y^2 h_{\mu \nu} \left([\Pi] \eta^{\mu\nu}- \Pi^{\mu \nu} \right)+ \dots
\ee
After diagonalization $\pi$ has the kinetic term
\be
\( \pa_y^2 \partial_\mu \pi \) \partial^\mu \pi.
\ee
From a four-dimensional point of view, we should canonically normalize $\pi$ in the sense $\pi\rightarrow (1/\pa_y) \pi$. Thus the non-local `canonical normalization' $1/\pa_y$ introduces a $\pa_y$ dependence in the strong coupling scale. Since $\pa_y$ can be made arbitrarily small, we can decrease the apparent strong coupling scale of its interactions. It is precisely this non-local normalization that is reflected in the low strong coupling scale of the discretized theory. However the continuum theory the ``scale" $1/\partial_y$ is a gauge artifact; we may always move to another gauge, such as de Donder gauge, where this issue never arises. In the discretized theory, as we have seen, the low strong coupling scale is physical, and so the process of discretization has made the gauge artifact into a physical scale.

We can see the origin of the dangerous interactions even more clearly by using the shift as our fundamental variable in the continuum theory. Writing
\be
K_{\mu\nu} = \frac{1}{2}\left(\partial_y g_{\mu\nu} - \nabla_\mu N_\nu - \nabla_\mu N_\nu \right)\,,
\ee
and focusing on the scalar part $N_\mu = -\nabla_\mu \phi$ we have
\ba
\mathcal{L}_m &\propto & \left([K^2]-[K]^2\right) \nn \\
&=& \Big[  (\nabla_\mu \nabla_\nu \phi)^2 - (\square \phi)^2 + \partial_y g_{\mu \nu}\( \nabla^\mu \nabla^\nu \phi-g^{\mu\nu}\Box \phi\) \\
&&\phantom{m^2 \mpl^2 \Big[} - \frac 14\((\partial_y g_{\mu\nu}) (\partial_y g^{\mu\nu})+(g^{\mu\nu}\p_y g\mn)^2\)\Big]\nn\,.
\ea
We see, as before, that the scalar part gets its kinetic term from mixing with $\pa_y g_{\mu\nu}$ and gets the canonical normalization $1/\pa_y$. The seemingly  dangerous interactions between $\phi$ and the metric then come from $(\nabla \nabla \phi)^2$ pieces. Up to a total derivative we have
\be
\nabla_\mu \nabla_\nu \phi \nabla^\mu \nabla^\nu \phi - (\square \phi)^2 = - R^{\mu\nu}\nabla_\mu \phi \nabla_\nu \phi
\ee
If we linearize around Minkowski, this becomes a $\partial^2 h (\partial \phi)^2$-type of interaction, precisely one of the dangerous interactions that we had found in the discretized theory. Thus in the continuum theory we apparently have the same strong coupling issue. Indeed even classical perturbation theory will appear strongly coupled. However we know that in the continuum theory the apparent strong coupling can be resolved by choosing a different gauge.

\subsubsection{Keeping the Lapse}

In the continuum theory, the strong coupling is fake. This is most easily seen in a gauge with $\N\ne 1$.
More precisely, we have
\be
S_{\rm GR, 5d} = \frac{M_5^3}{2} \int \d^5 x \sqrt{-g} \left(\N R - \frac{1}{\N}\left([K^2] - [K]^2 \right) \right)\,.
\ee
Performing a conformal transformation $g=\N^{-1} \tilde g$ this becomes
\be
S_{\rm GR, 5d} = \frac{M_5^3}{2} \int \d^5 x \sqrt{-\tilde g} \left(\tilde R - \frac{3}{2}(\p_\mu \N)^2 - \N^{-3}\left([K[g]^2] - [K[g]]^2 \right) \right)\,.
\ee
Now we linearize around Minkowski, including fluctuations in the shift, $\N=1+\delta \N/M_5$. We see that
\be
S_{\rm GR, 5d}  \supset \int \d^5 x\  -\frac{1}{4} h^{\mu\nu}\mathcal{E}_{\mu\nu\rho\sigma}h^{\rho\sigma} - \frac{3}{4} \left(\p_\mu \delta \N \right)^2 \,.
\ee
Crucially, we do not need to \stu to find a kinetic term for the helicity zero mode which is now propagating within $\delta \N$. Thus the canonical normalization of the helicity zero mode does not come with the large factor $1/\pa_y$. This shows that there are no interactions at the scale $\Lambda$, and since no field is canonically normalized in a way which depends on involves negative powers of $\p_y$, there cannot be any operator that arises at scale which depends on $\p_y$ (which in the discretized counterpart would correspond to a dependence on the IR scale $N$).

What this demonstrates is that we could never have hoped to recover general relativity in the discretized theory without including the lapse with its own dynamics which allows for the helicity-0 mode to have  a kinetic term even in the low KK mode limit, $\p_y \to 0$. If we freeze the lapse, already at the level of the continuum theory we see appearing the same strong coupling operators (which in the continuum case can be removed by  an appropriate shift of the lapse, but this involves putting some dynamics back in the lapse). Thus by discretizing without the lapse, we are giving up the ability to get rid of the low strong coupling scale.

We conjecture that a discretization procedure that keeps the lapse will have a well behaved continuum limit. For instance, working in five dimensional de Donder gauge requires a nonzero lapse.
By introducing the lapse the theory will no longer be described by a ghost-free multi-gravity Lagrangian, and we expect there to be a new degree of freedom beyond the five of five-dimensional General Relativity. While it maintains $y$ diffeomorphism symmetry at the quadratic level, making it gauge equivalent to Fierz-Pauli, at the nonlinear level the $y$ diffs must be broken, and the resulting formulation contains new degrees of freedom.

Thus to check that the continuum limit is recovered, one must study the canonical normalization, mass, and interactions of the new degrees of freedom. This is clearly an interesting question to study, however such a detailed calculation is beyond the scope of the present work. We emphasize that (1) we have shown that if it is possible to have a well-behaved continuum limit at all, it must involve the lapse, and (2) unlike in the case with $\N=1$, there is nothing in the continuum theory to suggest the existence of a low strong coupling scale in the discretized theory.

Thus we are faced with different alternatives when we discretize General Relativity in the vielbein language\footnote{Discretizing in the metric breaks unitarity at an undesirably low scale.},
\begin{itemize}
\item Either we fix all gauges, satisfy the phase-space constraints first and discretize the theory afterwards. This leads to a discretized theory which does not have any ghost, but breaks Lorentz invariance at the scale of the highest mode. Thus the resulting truncated theory breaks Lorentz invariance.
\item As we have seen, an alternative is to fix a gauge where the lapse is frozen $\N=1$.  The resulting truncated theory then maintains Lorentz invariance and $5N$ degrees of freedom, but becomes strongly coupled at a low energy scale $\Lambda_c$. This strong coupling is a manifestation of the Vainshtein mechanism which is a welcome feature from a four-dimensional perspective.
\item Finally, the only hope to discretize without breaking Lorentz invariance nor introducing strong coupling at a low scale is to maintain the lapse as dynamical. In the truncated theory, this procedure will then most likely break unitarity at best at the scale of the highest KK mode.
\end{itemize}
If we want to build a deconstruction framework that can recover General Relativity in the continuum limit, then it would be important to not introduce a physical strong coupling scale $\Lambda_c$. This scale controls the onset of the Vainshtein regime, but General Relativity does not have a Vainshtein effect. On the other hand we would be ready to accept new degrees of freedom at the scale $m$, the highest Kaluza-Klein mass in the theory, because we would be willing to consider the discretized theory to be an effective field theory with a cutoff given by $m$.

However from the point of view of constructing massive gravity and multi-gravity we can live with a low strong coupling scale. What we cannot accept are new degrees of freedom beyond the $5N$, because we are ultimately interested in the cases where $m$ is small. This leaves us with a low strong coupling scale $\Lambda_c$ which determines when the Vainshtein mechanism begins to become important. \textbf{We emphasize that the low strong coupling scale, $\Lambda_c$, is not necessarily the cutoff of an effective theory.} Instead, the Vainshtein mechanism provides for the possibility that the theory is simply UV complete, at least up to the Planck scale which may be understood either through a dual formulation \cite{deRham:2013hsa,Curtright:2012gx,Gabadadze:2012sm} or otherwise \cite{Dvali:2010jz,Alberte:2012is}.
In fact, since the Vainshtein mechanism is tied to the strong coupling scale, the low strong coupling scale is actually very important for the observational relevance of multi-gravity theories.

\section{Discussion}
\label{sec:Discussion}
We have established a connection between massive gravity, bigravity and multi-gravity in four dimensions and standard General Relativity in five dimensions. All of these results extend to the general dimension case. The special structure of the dRGT mass terms which removes the Boulware-Deser ghost can now be seen as deriving from the ghost free properties of General Relativity itself. Five dimensional General Relativity compactified on a circle of arbitrary radius is equivalent to a consistent low energy effective field theory of one massless graviton, one massless scalar, one massless gauge field and $N-1$ massive gravitons with the mass terms given by the multi-gravity generalizations of the dRGT mass terms. Fundamentally these results are not surprising since they are consistent with the normal Kaluza-Klein framework. However it is only due to the recent developments in massive gravity that we now know how to write down a closed form expression for this statement using the deconstruction method. At the linearized level the deconstruction and Kaluza-Klein frameworks are just discrete Fourier transforms of each other. At the nonlinear level in terms of the metric they are extremely complicated field redefinition away from each other. However in terms of the vierbein they remain as discrete Fourier transforms meaning that they encode all of the same physics.

Consistently with previous work, we have shown that there is a low strong coupling scale present in the discrete theory that remains even when the mass of the highest Kaluza-Klein mode is pushed to the five-dimensional Planck scale. This strong coupling
scale is related to the poor behaviour of the gauge $\N=1$ for low KK momenta modes in the continuum theory. Picking a different gauge where the lapse remains dynamical resolves the poor behaviour in the continuum theory. We conjecture that a discretization prescription that keeps the lapse would prevent the emergence of a low strong coupling scale and have a smooth continuum limit at the price of introducing new degrees of freedom (since the resulting theory is not equivalent to a ghost-free multi-gravity theory). These results represent a tension faced in discretization between avoiding a low strong coupling scale and preventing the onset of new degrees of freedom.

The failure of multigravity theories to recover general relativity in the continuum limit should not be taken as a failure of the multi-gravity theories, but rather as an expression of the fact that multi-gravity theories are a solution to a different problem. Multi-gravity theories are meant to make sense as four dimensional theories. They rely on a low strong coupling scale in order to have a Vainshtein mechanism, this is how they are able to recover four-dimensional General Relativity. However, precisely this fact also prevents them from ever looking like a truly five-dimensional theory.

Several extensions of these results are possible. One can construct more general multi-gravity theories by discretizing General Relativity with multiple extra dimensions. One may consider higher order (Lovelock) terms living in the extra dimensions beyond the Einstein Hilbert term \cite{Progress}, and see what their discretized limit and whether additional ghost free interactions, such as those conjectured in \cite{Hinterbichler:2013eza}, can be found. Finally the deconstruction framework suggests that it is natural for matter to couple to different metrics, at least at zeroth order in derivatives. The full implications of this have not been taken into account in most discussions up to now where it is assumed that all matter in the Standard Model couples to just a single metric. We leave these interesting questions to future work.

\acknowledgments

We would like to thank Cedric Deffayet, Kurt Hinterbichler and Nick Ondo for useful discussions.
CdR is supported by Department of Energy grant DE-SC0009946.
AJT is supported by a Department of Energy Early Career Award. AAM is supported by an NSF GRFP fellowship.

%%%%%%%%%%%%%%%%%%%%%%%%%%%%%%%%%%%%%%%%%%%%%%%%%%%%%%%%%%%%%%%%%%%%%
%%%% Bibliography

%\newpage

  \end{document}